\documentclass[english,12pt]{iopart}
\usepackage[T1]{fontenc}
\usepackage[latin9]{inputenc}
\usepackage{textcomp}
\usepackage{amstext}
\usepackage{graphicx}
\usepackage{esint}

\makeatletter

\newcommand{\lyxmathsym}[1]{\ifmmode\begingroup\def\b@ld{bold}
  \text{\ifx\math@version\b@ld\bfseries\fi#1}\endgroup\else#1\fi}

\usepackage{iopams}
\usepackage{setstack}

\makeatother

\usepackage{babel}
\begin{document}

\title{Simulating Bell violations without quantum computers}

\author{P D Drummond, B Opanchuk, and L Rosales-Z\'arate and M D Reid}

\address{Centre for Atom Optics and Ultrafast Spectroscopy, Swinburne University
of Technology, Melbourne 3122, Australia}
\begin{abstract}
We demonstrate that it is possible to simulate Bell violations using
probabilistic methods. A quantum state corresponding to optical experiments
that violate the Bell inequality is generated, demonstrating that
these quantum paradoxes can indeed be simulated probabilistically.
This provides an explicit counter-example to Feynman's claim that
such classical simulations could not be carried out.
\end{abstract}

\submitto{\PS}

\maketitle

\section{Introduction}

The simulation of quantum dynamics is a hard problem in physics for
systems with many degrees of freedom. While large classical systems
can be readily simulated with computers, this is difficult with large
quantum systems due to the exponentially growing size of Hilbert space.
Since an expansion in eigenstates of the Hamiltonian is virtually
impossible for many-body systems, one possible solution is to use
probabilistic sampling. This approach was apparently ruled out by
a claim of Feynman~\cite{Feynman1982}, where he asked

\begin{quote}

\textbf{\emph{Can quantum systems be probabilistically simulated by
a classical computer}}\emph{?}

\end{quote}

His answer to the question was:

\begin{quote}

\textbf{\emph{If you take the computer to be the classical kind and
there's no changes in any laws, and there's no hocus-pocus, the answer
is certainly, No!}}

\end{quote}

This led Feynman to propose the use of quantum computers for these
types of simulation, and his argument has motivated extensive research
on quantum computing. However, large-scale quantum computers are not
yet available. An important question therefore is whether probabilistic
simulation with classical computers are truly excluded, as this claim
would certainly imply.  Is there some way that one can in fact use
probabilistic sampling with existing digital computers for these challenging
tasks? And if so, to what degree of practicality?

Given the known decoherence problems in constructing quantum computing
hardware for quantum simulations, practical alternative strategies
using software would be extremely useful. Here we demonstrate, by
carrying out a simulation, that Bell's theorem does not rule out probabilistic
simulation methods. Our results treat the case considered by Feynman,
which is a four mode photonic state used in experimental demonstrations,
equivalent to two correlated spin-half particles. 

This allows us to demonstrate, by means of explicit, direct computer
simulation, that the answer to Feynman's original question is actually
\textit{\emph{``}}\textbf{\textit{Yes}}\textit{\emph{''}}. Our main
emphasis is the fundamental question of whether it is possible to
probabilistically treat Bell violations, which was claimed to be impossible
in this early literature. In fact, the ``not so black and white''
nature of Feynman's claim was always apparent, given the existence
of positive phase space methods that were known at the time. It was
sometimes interpreted that these methods would only be useful for
semiclassical states, rather than the Bell state. We show this is
not so by carrying out localized probabilistic sampling of a Bell-violating
quantum state for the first time, using random sampling to obtain
results in complete agreement with quantum mechanics.

One may ask then, in Feynman's words: what is the ``\emph{hocus-pocus}''
that allows these simulations? Very simply, we only require that average
correlations of the simulation outputs correspond to quantum mechanics.
We do not ask the measurement values and their probabilities to be
exactly identical to quantum mechanics. This allows us to average
over probabilities of phase-space variables whose values are not integer
eigenvalues. Another possible way to overcome this argument could
be to treat nonlocal simulation, which exploits another loophole in
Feynman's logic: a simulation doesn't have to satisfy locality. We
do not investigate this option here, since our simulations use local
variables, although removing this restriction may provide additional
resources.

We use positive phase-space distributions of quantum mechanics for
our purpose~\cite{Husimi1940,Drummond1980}. These exist for every
quantum state, including Bell states. Their statistical moments correspond
to quantum correlations. The advantage of probabilistic sampling \textemdash{}
which Feynman also realized \textemdash{} is that the exponential
growth of memory size with the number of qubits is removed. This potentially
eliminates the problem of rapid increase in memory size with orthogonal
expansions. The details of the important related issues of computational
efficiency, error propagation and general scalability cannot be treated
here for space reasons. 

Instead, we simply wish to examine Feynman's original argument, by
asking if probabilistic simulations of quantum states can be carried
out along the lines proposed in his original paper, i.e., by constructing
correlations using products of variables obtained from probabilistic
sampling.

\section{Feynman's argument}

Earlier investigations on the limits to computation had focused on
the dissipation of energy in standard logic operations. Accordingly,
it was rather natural to investigate the possibility first of dissipationless
logic~\cite{Bennett2000}, then of quantum logic~\cite{Feynman1982}
as alternatives.

Feynman's paper addressed the issue of exponential scaling in quantum
dynamics. Was this a problem that classical computers could be used
to solve? If not, perhaps quantum computers would be needed. Feynman's
logic here was reasonably clear. Random sampling methods are known
as a way to treat many other exponentially complex problems. If one
could rule out probabilistic sampling by considering a known quantum
state, this entire strategy could then be eliminated for quantum simulations. 

To prove this, Feynman turned to Bell's theorem~\cite{Bell1964},
which shows that all local hidden variable (LHV) theories are inequivalent
to quantum mechanics. His argument used a small quantum system without
inherent exponential complexity. However, by assuming that a probabilistic
simulation is equivalent to a hidden variable theory, Feynman could
argue from this simple case that no general probabilistic method was
possible. It is this claim that we investigate here.

To explain this in greater detail, we recall that a Bell inequality
is a constraint on observable correlations of a physical system that
obeys a local hidden variable theory. This is a theory that has the
property of local realism, as defined by Einstein. In the case of
particles emitted from a common source, measurements of two spatially
separated observers are obtained by taking random samples of a common
parameter $\lambda$. 

Measured values are then functions of some local detector settings
and the hidden parameter $\lambda$, which could be any variable.
Mathematically, the correlations in a hidden variable theory are obtained
from a probabilistic calculation of form:

\begin{equation}
\left\langle \hat{X}_{1}\hat{X}_{2}\right\rangle =\int X_{1}(\lambda)X_{2}(\lambda)P(\lambda)d\lambda\,.\label{eq:LHV}
\end{equation}
Here, $X_{j}(\lambda)$ have values that correspond to the experimentally
measured eigenvalues. This is just how one might expect to carry out
a probabilistic simulation. However, as we will see, this restriction
on the values of $X_{j}(\lambda)$ \textemdash{} which is needed for
LHV theories \textemdash{} is not essential for a probabilistic simulation.

\section{Optical Bell violations}

A popular route to Bell violation experiments is using an atomic cascade,
or more recently using parametric down-conversion, with the resulting
state: 
\begin{equation}
\vert\Psi_{B}\rangle=\frac{1}{\sqrt{2}}\left(a_{1+}^{\dagger}a_{2+}^{\dagger}+a_{1-}^{\dagger}a_{2-}^{\dagger}\right)\vert0\rangle.\label{eq: Bell state}
\end{equation}

Here we suppose that $a_{1\pm}^{\dagger}$ creates a photon in spatial
mode 1, which is detected at site $A$ with polarization of $s=\pm1$
respectively, and similarly for operators of mode 2 detected at site
$B$. The version of Bell's inequality given by Clauser, Horne, Shimony
and Holt, (the CHSH form)~\cite{Clauser1969} is especially important,
as it gives LHV limits to the expected correlation for the above experiment
conducted by Alice and Bob:

\begin{equation}
\mathbf{C}[A,B]+\mathbf{C}[A,B^{\prime}]+\mathbf{C}[A^{\prime},B]-\mathbf{C}[A^{\prime},B^{\prime}]\leq2,
\end{equation}
where $\mathbf{C}[A,B]$ is the correlation, $A$ and $A^{\prime}$
are measurements at location A with two different polarizer angles,
while $B$ and $B^{\prime}$ are the corresponding measurements at
location B. It is usually assumed that assume the observed values
are $+1$ or $\lyxmathsym{\textminus}1$. We note that a calculation
within quantum mechanics shows that, for a singlet quantum state known
as the Bell state, the Bell inequality is predicted to be violated.
Quantum mechanics predicts that at a relative polarizer angle $\theta=\pi/8$:

\begin{eqnarray}
\Delta(\theta)=\frac{1}{2}\left[\langle\hat{A}\hat{B}\rangle+\langle\hat{A}^{\prime}\hat{B}\rangle+\langle\hat{A}\hat{B}^{\prime}\rangle-\langle\hat{A}^{\prime}\hat{B}^{\prime}\rangle\right]-1 & = & \sqrt{2}-1>0.
\end{eqnarray}

Feynman's claim was that a classical probabilistic simulation could
not replicate this violation. Hence he argued that classical probabilistic
methods using a computer could not be used to simulate quantum dynamics.

\section{Positive P-representation}

To demonstrate a counter-example - a probabilistic simulation of the
bipartite Bell inequalities given above - we use the positive-P representation~\cite{Drummond1980}.
This is an expansion of an arbitrary density matrix $\widehat{\rho}$
in coherent state projectors:
\begin{equation}
\widehat{\rho}=\int P(\vec{\alpha},\vec{\beta})\widehat{\Lambda}(\vec{\alpha},\vec{\beta})d^{2M}\vec{\alpha}d^{2M}\vec{\beta}.
\end{equation}

Here, the projector is $\widehat{\Lambda}(\vec{\alpha},\vec{\beta})=\left|\vec{\alpha}\right\rangle \left\langle \vec{\beta}^{*}\right|/\langle\vec{\beta}^{*}\vert\vec{\alpha}\rangle$,
and $\left|\vec{\alpha}\right\rangle =\left|\alpha_{1}.\ldots\alpha_{n}\right\rangle $
is a multi-mode coherent state of a bosonic field. The probability
function $P(\vec{\alpha},\vec{\beta})$ is defined on an\textbf{ }enlarged,
nonclassical phase-space, which allows positive probabilities. This
method leads to an exact mapping between the quantum mechanics of
any bosonic field, and a phase-space probability distribution \cite{Drummond1980}.
This was already known by 1982.

The correlations of quantum counts $\hat{n}_{i}=\hat{a}_{i}^{\dagger}\hat{a}_{i}$
at different locations are simulated using: 
\begin{equation}
\left\langle \hat{n}_{i}\ldots\hat{n}_{j}\right\rangle =\int n_{i}\ldots n_{j}P(\vec{\alpha},\vec{\beta})d^{2M}\vec{\alpha}d^{2M}\vec{\beta}.\label{+Pcorrels}
\end{equation}
where $n_{i}\equiv\alpha_{i}\beta_{i}$. The effects of a polarizer
are simply obtained on taking linear combinations of mode amplitudes,
just as in classical theory.

One can obtain the positive-P distribution using a variety of methods,
since the representation is not unique. Here, we represent the photonic
Bell state of Eq (\ref{eq: Bell state}) using a generic construction
which exists for all quantum states \cite{Drummond1980,Drummond1983}:

\begin{equation}
P(\vec{\alpha},\vec{\beta})=\frac{1}{\left(2\pi\right)^{2M}}e^{-\left|\vec{\alpha}-\vec{\beta}^{*}\right|^{2}/4}\left\langle \frac{\vec{\alpha}+\vec{\beta}^{*}}{2}\right|\widehat{\rho}\left|\frac{\vec{\alpha}+\vec{\beta}^{*}}{2}\right\rangle .\label{eq:P-from-rho}
\end{equation}

There is a remarkable similarity between the hidden variable theory~(\ref{eq:LHV})
of Bell, and the positive-P formula~(\ref{+Pcorrels}) for quantum
correlations. However, while a hidden variable theory obeys Bell's
theorem, the positive-P theory is fully equivalent to quantum mechanics,
and can violate Bell inequalities. The reason for the difference is
due to the different quantities calculated in the correlations. The
fundamental observables in Bell's case, of form $X(\lambda)$, are
equal to observed integer photon counts. 

The corresponding observables in the positive-P case, of form $n\left(\vec{\alpha},\vec{\beta}\right)$,
are complex numbers whose mean values and correlations correspond
to physical means and correlations. This difference allows the positive-P
distribution to be equivalent to quantum mechanics, even though it
looks similar to a hidden variable theory. This point was made in
an article by Reid and Walls~\cite{ReidWalls1986}, which proposed
the modern Bell inequality experiments using parametric down-conversion
to generate photon pairs.

\section{Simulation results}

In our simulations, we choose to expand the Bell state, Eq~(\ref{eq: Bell state})
using the standard method of Eq~(\ref{eq:P-from-rho}). While it
is also possible to solve the dynamical stochastic equations for the
parametric amplifier used in experiments \cite{McNeilGardiner1983},
here we are simply interested in demonstrating that probabilistic
sampling of the Bell state is possible. In the ideal Bell case, the
actual distribution has $M=4$ modes with $16$ real dimensions, having
the form:

\begin{equation}
P(\vec{\alpha})=\left(\frac{\left|\vec{\alpha}_{A+}\cdot\vec{\alpha}_{B+}\right|^{2N}}{\pi^{8}\left(N+1\right)\left(N!\right)^{2}}\right)e^{-\left|\vec{\alpha}_{+}\right|^{2}-\left|\vec{\alpha}_{-}\right|^{2}}\,.
\end{equation}

Here $\vec{\alpha}_{\pm}=\left(\vec{\alpha}\pm\vec{\beta}^{*}\right)/2$
are the sum and difference coordinates respectively. Clearly, $\vec{\alpha}_{-}$
can be sampled using Gaussian variates. The notation $\vec{\alpha}_{A,B+}$
indicates the coherent state sum vector projected on the $A$ and
$B$ observers respectively. The sum distribution is more complex,
but can be readily sampled using the von Neumann rejection method,
with a standard lambda distribution as the reference distribution. 

\begin{figure}
\centering{}\includegraphics{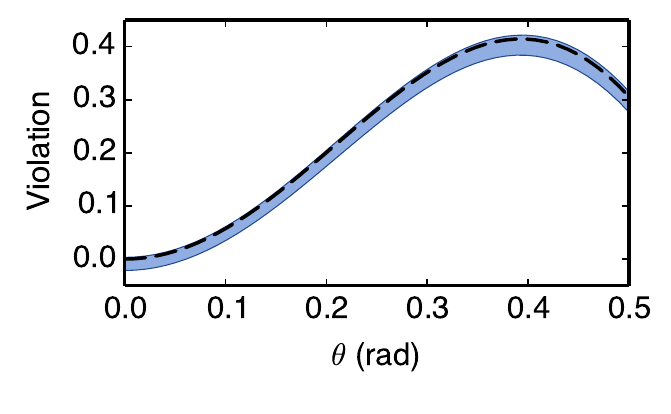}\caption{Probabilistic violation of a Bell inequality with $2\times10^{6}$
random samples. The simulated Bell violation $\Delta(\theta)$ is
graphed as a function of the relative polarizer angle $\theta$. The
filled area corresponds to the estimated error range around the mean
of $\Delta(\theta)$ for the sampled state, while the dashed line
is the quantum mechanical prediction.\label{Bell-violation}}
\end{figure}

\begin{figure}
\centering{}\includegraphics{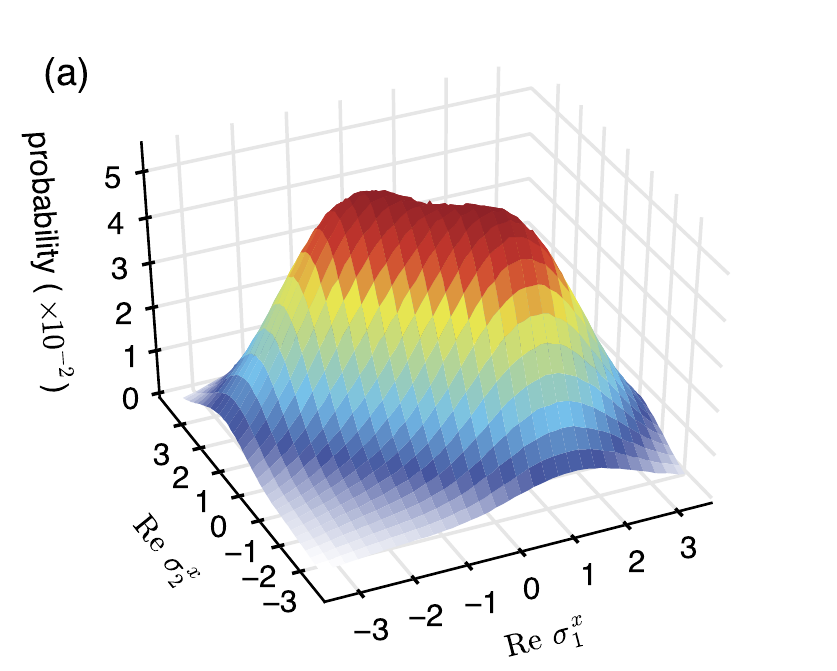}\includegraphics{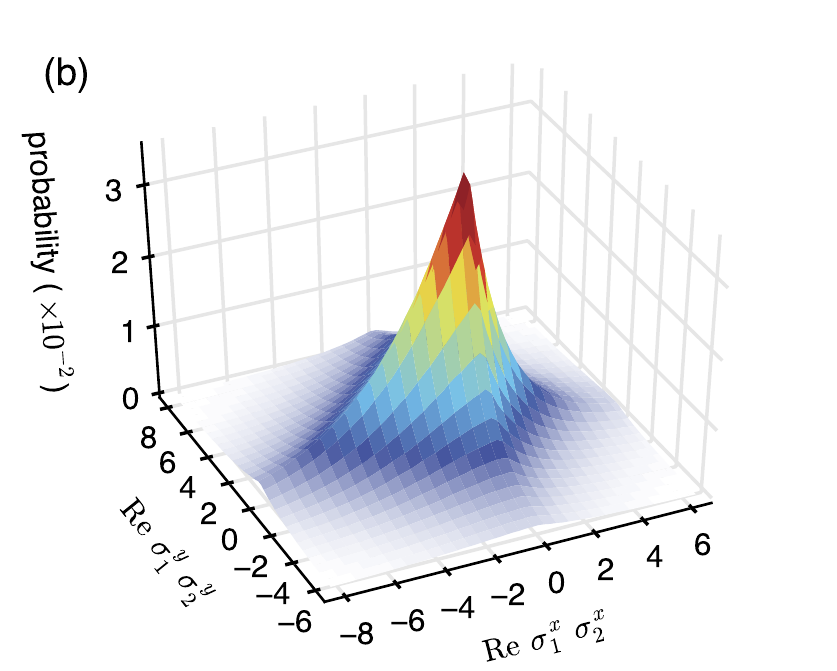}\caption{Distribution of variables correspond to spins and correlations.\label{Distribution}}
\end{figure}

The resulting Bell violation is graphed in Fig~(\ref{Bell-violation}).
This  demonstrates complete agreement with quantum predictions, up
to a sampling error which can be reduced at will by taking more samples.
However, the results of Fig~(\ref{Distribution}) are more interesting.
This figure shows the joint distribution of the Schwinger variables
that correspond to the spin projections and their correlations. In
a physical measurement, these would all have eigenvalues of $\pm1$.
Instead, we see that the variables corresponding to spin measurements
go outside their quantum bounds. Intriguingly, this is exactly what
is predicted for weak measurements, which suggests that a close relationship
exists between these simulations and the concept of a weak measurement.

\section{Conclusions}

Our main result is that the earliest argument leading to quantum computers
is wrong. There is no impediment to simulating Bell violations probabilistically.
Our phase-space simulations correctly generate the means and correlations
that are predicted by quantum mechanics. Such probabilistic simulation
methods have already allowed simulations of the quantum dynamics of
quantum solitons~\cite{Carter:1987,Corney:2006_ManyBodyQD} and colliding
Bose-Einstein condensates~\cite{Deuar2007}, with up to $10^{6}$
modes and $10^{5}$ particles. The growth of sampling error and other
scaling issues are important limitations, and will be treated elsewhere.

Quantum logic based encryption~\cite{BennettCrypto} and atomic clocks~\cite{WinelandClock}
are already successful. To develop such technologies in future \cite{YamamotoPhysRevX.2.031007},
understanding the consequences of non-ideal behavior is very important.
Probabilistic algorithms could therefore have an application to the
design of these devices. Importantly, fundamental tests of quantum
mechanics require simulation methods that do not depend on quantum
theory being correct, which is a strong reason to investigate these
issues further.

\ack{}{}

L. E. C. R. Z. acknowledges financial support from CONACYT, Mexico.
P. D. D. and M. D. R. acknowledge Australian Research Council funding
of a Discovery grant.

\section*{References}{}

\bibliographystyle{iopart-num}
\bibliography{BellQLsims}

\end{document}